\documentclass[a4paper, 12pt]{article}
\usepackage{graphicx}
\usepackage{epsf}
\begin{document}
\date{}
\title{Helical inward orbits of high-$Z$ impurities}
\author{F. Spineanu and M. Vlad \\
National Institute of Laser, Plasma and Radiation Physics\\
Magurele, Bucharest 077125, Romania}
\maketitle

\begin{abstract}
In a tokamak the edge and the central region may become connected by helical orbits resulting from the
combination of global poloidal rotation and localized radial shifts. The rotation sustained by the Stringer mechanism 
and the baroclinic effect are able to generate such helical orbit.  This can be a contribution to the inward flow and central accumulation 
of high-$Z$ impurity ions.
\end{abstract}

The high $Z$ impurity ions that enter in the tokamak plasma from the first
wall are a source of energy loss via radiation and can be a danger for the
reactor regime. The experiments show that the density of heavy ion impurity
contamination has a substantial poloidal nonuniformity, with maximum in the
Low Field Side, and that in the $H$ mode there is an accumulation of ions in
the center of the discharge. The high-$Z$ ions have a fast displacement
toward the plasma core (possibly faster than diffusion). 

The physical picture of the access and the accumulation of heavy impurity
ions in the center of the discharge is rather complex and have received
several explanations.

There is a collisional momentum transfer from the background ions to the
high-$Z$ ion. The equilibrium gradient of the background ions, with decrease
of density from the center toward the edge, may suggest that to an impurity
ion it is collisionally imposed an outward diffusion. However, the argument
of {Longmire and Rosenbluth \cite{LR56} }shows that the impurity ions
actually have opposite diffusion, \emph{i.e.} radially inward
("counter-gradient"). In an alternative picture, the impurity ions must have
inward diffusion in order to preserve ambipolarity: the background ions have
outward diffusion (fixed by the gradient of their density) while the
electrons, that are tied to the magnetic surface, cannot follow. This would
produce a non-ambipolar radial flux and the impurities are requested to
respond by an inflow that compensate for the radial outflow of the basic
ions, $\Gamma _{i}-\Gamma _{imp}\times Z=0$. In the presence of ICRH, the
minority with large perpendicular energy have large banana orbits and their
ourward half traverses a small region at the outboard edge. It then results
an accumulation of charge in this area with generation of an electric field.
The poloidal projection of this electric field, combined with the confining
magnetic field, generates a radially inward flow of heavy ions.

There are mechanisms that can contribute to the faster-than-diffusion access
and accumulation of heavy ions in the center, like the Turbulence
Equipartition density pinch and the transient ballistic flows when the
discharge is divided into a toroidal dipolar symmetric vortex.

\bigskip 

We here examine a new possiblity: the existence of radially inward helical
orbits connecting the plasma edge with the center. 

The basic elements that make possible an inward spiraling motion for an
element of plasma, are a global poloidal rotation of plasma and the \emph{%
baroclinic} effect that produces local radial deviations from a purely
poloidal orbit.

For the poloidal rotation we consider the mechanism of Stringer, able to
produce a poloidal torque when there is a poloidal nonuniform radial flux of
particles. 

\bigskip 

The Stringer spontaneous (\emph{i.e. }without externally applied torque)
rotation results from the conservation laws \cite{hassamdrake}, \cite%
{hassamprl}, \cite{hazeltineleerosenbluth}. The equation for the plasma
poloidal velocity is 
\begin{equation}
\frac{\varepsilon }{q}\left( 1+2q^{2}\right) \left( \frac{\partial V_{\theta
}}{\partial t}+\gamma _{MP}V_{\theta }\right) +qV_{\theta }\frac{1}{nr}\frac{%
\partial }{\partial r}\left( nr\left\langle 2\cos \theta
\;v_{r}\right\rangle \right) =0  \label{eq7}
\end{equation}%
Here $\gamma _{MP}$ is the rate of damping of poloidal rotation by the
magnetic pumping \cite{suyush}. The radial flux is $\Gamma _{r}\approx
\left\langle 2\cos \theta \ n\widetilde{v}_{r}\left( \theta \right)
\right\rangle $ where the surface averaging operation is $\left\langle
f\right\rangle \equiv \oint \frac{d\theta }{2\pi }\left( 1+\varepsilon \cos
\theta \right) \;f$. The Stringer mechanism is a basic neoclassical effect.
The variation on surface of the density $\widetilde{n}$ and of the radial
diffusive flow $\widetilde{v}_{r}$ produces a local flux $\widetilde{\Gamma }%
_{r}\left( \theta \right) $ that, due to the tokamak geometry, has non-zero
divergence. A poloidal flow must exist to compensate and $V_{\theta }$ would
grow indefinitely (a shock line is eventually created at the inner side of
plasma), The magnetic pumping damping $\gamma _{MP}$ is very efficient in
suppressing the purely neoclassical Stringer torque. However if the radial
flux $\widetilde{\Gamma }\left( \theta \right) $ with variation on poloidal
angle $\theta $ is due to anomalous transport, the torque can be higher than 
$\gamma _{MP}$ and the plasma has poloidal rotation. This is a mechanism
independent to the Reynolds stress, which is usually considered the source
of poloidal rotation in the $H$ mode and in Internal Transport Barrier.
Several analyses on the Stringer effect induced by the poloidally
non-uniform anomalous transport fluxes \cite{hassamdrake}, \cite{hassamprl}
show that it can be much higher than the Reynolds stress-induced torque and
this permit us to consider in the following that the plasma has a global
poloidal rotation.

\bigskip 

\begin{figure}[h]
\begin{center}
\includegraphics[height=8cm]{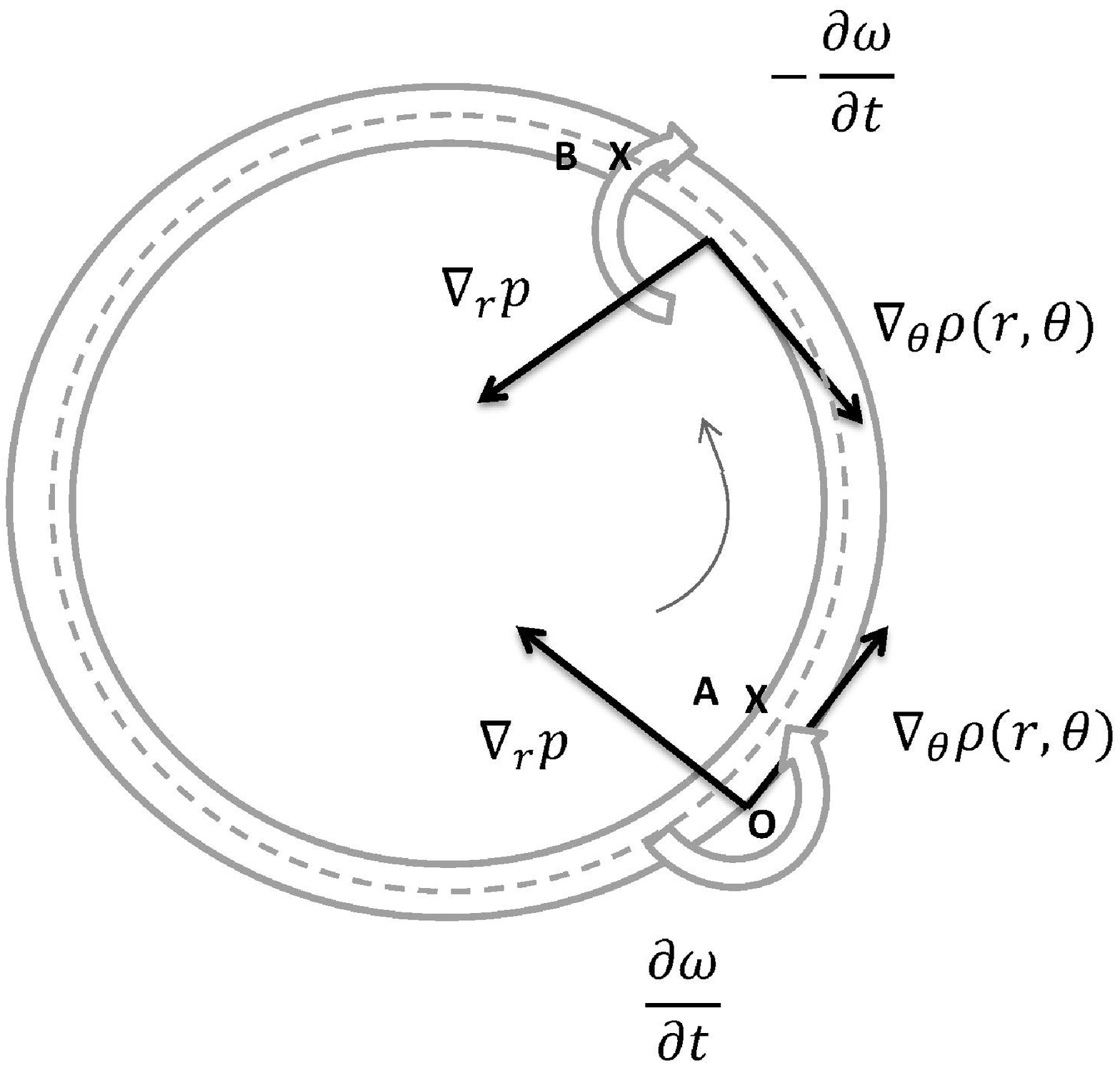}
\end{center}
\end{figure}

Now we discuss the effect of the poloidal nonuniformities of the density
from a different angle. We consider  $\widetilde{n}\left( r,\theta \right) $
the perturbation to the density $n_{0}\left( r\right) $ uniform on surface.
It has the maximum on the equatorial plane and is symmetric above and below
the plane, with identical spatial (\emph{i.e. }$\sim \theta $) decay. This
means that $\widetilde{n}\left( r,\theta \right) $ decreases on each surface 
$r$, from the maximum on the equator, along the arc distances $\pm
r\left\vert \theta \right\vert $. The particular function is not essential
now and we only remark that there are gradiants $\mathbf{\nabla }_{\theta }%
\widetilde{n}\left( r,\theta \right) $ tangent to the surface $r$ with
orientation $-\widehat{\mathbf{e}}_{\theta }$ above and $\widehat{\mathbf{e}}%
_{\theta }$ below the plane ($\widehat{\mathbf{e}}_{\theta }=r\mathbf{\nabla 
}_{\theta }\theta $ is the versor of $\theta $-variation). Such gradients
induce a \emph{baroclinic} effect. For simplification we consider the
momentum balance equation and apply the $\mathit{curl}$ operator. 
\begin{equation}
\frac{\partial }{\partial t}\mathbf{\omega }+\mathbf{\nabla \times }\left[
\left( \mathbf{v\cdot \nabla }\right) \mathbf{v}\right] =\frac{1}{\rho ^{2}}%
\mathbf{\nabla }\rho \times \mathbf{\nabla }p-\gamma _{\omega }\omega 
\widehat{\mathbf{e}}_{\varphi }  \label{eq9}
\end{equation}%
where $\gamma _{\omega }$ is a schematic representation of any dynamical
damping of the vorticity, \emph{i.e.} limiting the development of higher
shear of the velocity. Projecting along the toroidal direction $\widehat{%
\mathbf{e}}_{\varphi }$, a stationary state of plasma rotation induced by
only this term should be 
\begin{equation}
\omega \sim \frac{c_{s}^{2}}{\gamma _{\omega }}\frac{1}{n_{0}}\frac{\partial
\ln \widetilde{n}}{r\partial \theta }\frac{\partial }{\partial r}\ln n_{0}
\label{eq10}
\end{equation}%
but we note that the gradients of $\widetilde{n}\left( r,\theta \right) $
have opposite sign above and below the equatorial plane and the effect is
cancelled. We are interested in the effect of the baroclinic term for an
element of plasma that, in the global poloidal rotation, traverses the two
regions where the gradients of $\widetilde{n}\left( r,\theta \right) $ are
substantial. The baroclinic effect is a source of vorticity 
\[
\frac{\partial \omega }{\partial t}\widehat{\mathbf{e}}_{\varphi }\sim
c_{s}^{2}\frac{1}{n_{0}^{2}}\frac{\partial \widetilde{n}\left( \theta
\right) }{r\partial \theta }\widehat{\mathbf{e}}_{\theta }\times \frac{%
\partial n_{0}}{\partial r}\widehat{\mathbf{e}}_{r}
\]%
($\widehat{\mathbf{e}}_{\varphi }=\widehat{\mathbf{e}}_{\theta }\times 
\widehat{\mathbf{e}}_{r}$) and this means the generation of a localized
effect of rotation. The approximate measure of the rotational effect which
acts for a short time interval $\Delta t$ is an angular displacement 
\[
{\mathrm angle}\sim \omega \Delta t\sim \frac{1}{n_{0}^{2}}\frac{\partial 
\widetilde{n}}{r\partial \theta }\frac{dn_{0}}{dr}\times \Delta t
\]%
In the Figure we represent the spatial deviation induced by this angular
displacement. An element of plasma initially located in $O$ is displaced, by
the baroclinic effect, radially inward to a new position $A$, at a smaller
radius $r_{A}<r_{O}$. It continues the global rotation and arrives in the
region where it is acted upon again by the baroclinic term but this time in
the opposite direction. However the two small, baroclinic-induced angular
displacements are not exactly equal and opposite since the element of plasma
is now at a radius $r_{A}$ which is different of the initial magnetic
surface $r_{O}$. The displacement will be to a new point $B$ which is not
exactly on the initial surface $r_{O}$ . A small effective radial
displacement results. At every turn in the global poloidal rotation, the
element of plasma receives a small radial displacement. Its orbit is
helically inward, a spiral that may connect the edge to the central region.
An estimation%
\begin{eqnarray*}
\delta r^{inward} &\sim &\delta r_{local}\left( r_{O}\right) -\delta
r_{local}\left( r_{A}\right)  \\
&\sim &\left[ \frac{L_{Z\theta }^{2}}{\left( r\Omega \right) ^{2}}\frac{1}{%
\rho _{Z}}\frac{dp_{0}}{dr}\right] \frac{\partial }{\partial r}\left[ \frac{%
L_{Z\theta }^{2}}{\left( r\Omega \right) ^{2}}\frac{1}{\rho _{Z}}\frac{dp_{0}%
}{dr}\right] 
\end{eqnarray*}%
where $L_{Z\theta }$ is the gradient length of the poloidal nonuniformity of
the density $\rho _{Z}$ of high-$Z$ impurities, $\Omega $ is the global
(Stringer) rotation of plasma, $p_{0}\left( r\right) $ is the equilibrium
plasma pressure. The condition for the existence of a helical orbit is $%
\frac{\partial }{\partial r}\frac{L_{Z\theta }}{V_{\theta }^{2}}>0$.
Inserting some physical data one obtains an order of magnitude of few
millimeters for the effective inward radial displacement at every poloidal turn and
a velocity of inward advancement of about $1\ \left( m/s\right) $.

\end{document}